# Correcting pervasive errors in RNA crystallography through enumerative structure prediction

Fang-Chieh Chou[1], Parin Sripakdeevong[2], Sergey M. Dibrov[3], Thomas Hermann[3] and Rhiju Das[1,2,4]*

[1]Department of Biochemistry, Stanford University, Stanford, CA 94305, USA

[2]Biophysics Program, Stanford University, Stanford, CA 94305, USA

[3]Department of Chemistry and Biochemistry, University of California at San Diego, 9500 Gilman Drive, La Jolla, CA 92093, USA

[4]Department of Physics, Stanford University, Stanford, CA 94305, USA

*Correspondence to: Rhiju Das. Phone: (650) 723-5976. Fax: (650) 723-6783. E-mail: rhiju@stanford.edu.

**Three-dimensional RNA models fitted into crystallographic density maps exhibit pervasive conformational ambiguities, geometric errors, and steric clashes. To address these problems, we present Enumerative Real-space Refinement ASsisted by Electron density under Rosetta (ERRASER), coupled to PHENIX (Python-based Hierarchical Environment for Integrated Xtallography) diffraction-based refinement. On 24 datasets, ERRASER automatically corrects the majority of MolProbity-assessed errors, improves average $R_{free}$ factor, resolves functionally important discrepancies in non-canonical structure, and refines low-resolution models to better match higher resolution models.**

Over the last decade, fruitful progress in RNA crystallography has revealed numerous 3D structures of functional RNAs, providing powerful information for understanding their biological functions[1, 2]. Nevertheless, RNA structures are typically solved at low resolution (typically >2.5 Å) compared to protein data. A recent report by the PDB X-ray Validation Task Force noted the ubiquity of bond geometry errors, anomalous sugar puckers, and backbone conformer ambiguities in RNA crystallographic models, and recommended that assessments of these features be included in PDB validation procedures[3]. There is thus a critical need for efficient algorithms that can resolve ambiguities in existing and future RNA crystallographic models.

The difficulty of resolving RNA crystallographic errors is underscored by limitations in currently available computational tools. RNABC (RNA Backbone Correction)[4] and RCrane (RNA Constructed using RotAmeric NuclEotides)[5] can identify and fix backbone conformer errors in some models. However, these methods anchor phosphates and bases to starting positions determined manually and thus only correct a subset of errors. Recent advances in Rosetta RNA *de novo* modeling[6-8] and electron-density-guided protein modeling[9, 10] have suggested that confident high-accuracy structure prediction may be feasible if guided by experimental data. We have therefore developed a method for Enumerative Real-space Refinement ASsisted by Electron density under Rosetta (ERRASER) and integrated it with the PHENIX tools for diffraction-guided refinement. The protocol is based on exhaustively sampling each nucleotide's possible conformations and scoring by the physically realistic Rosetta energy function supplemented with an electron-density-correlation score (see Methods and Supplementary Fig. 1). Based on a benchmark of published crystallographic datasets and newly solved RNA structures, we report that this automated pipeline resolves the majority of geometric errors while retaining or improving correlation to diffraction data.

To measure the effectiveness of the ERRASER-PHENIX pipeline, we collected a test set of 24 RNA-containing crystal structures deposited in the PDB, ranging from small pseudoknots to entire ribosomal subunits (Supplementary Table 1). In parallel, we tested the effectiveness of RNABC and RCrane as alternatives to the ERRASER refinement step in our protocol, as well as PHENIX alone. In the starting PDB-deposited structures, MolProbity tools revealed numerous potential errors in four classes: atom-atom steric clashes, high frequencies of outlier bond lengths or angles, 'non-rotameric' backbone conformations, and potentially incorrect sugar puckers[11]. While not all of these features are necessarily incorrect, their high frequencies in medium-to-low-resolution models (2.5-3.5 Å) compared to high-resolution models (< 2.0 Å) suggest that most are due to inaccurate fits[4, 5, 11, 12].

First, outlier bond lengths and angles (> 4 s.d. from reference values) in the crystallographic models have mean frequencies of 0.53 % and 1.18 % in the starting PDB coordinates. Some of these outliers are due to different ideal bond geometries used by different refinement packages, and thus PHENIX alone lowered the outlier frequencies substantially. Nevertheless, application of ERRASER-PHENIX gave greater improvement, eliminating all the outlier bond lengths and angles in the benchmark (Table 1 and Supplementary Table 2).

Second, ERRASER-PHENIX substantially reduced the steric clashes in RNA coordinates fitted into low-resolution electron density. In a bacteriophage prohead RNA test case (3R4F), the initial pervasive clashes were reduced by 80 % with ERRASER-PHENIX (Fig. 1a). Over the entire benchmark, the MolProbity clashscore (number of serious clashes per 1,000 atoms[11]) was reduced from an average of 18.0 to 7.0 (Fig. 2a). Other refinement approaches that use less stringent or no steric criteria gave higher average clashscores (Table 1 and Supplementary Table 3).

Third, a recent community-consensus analysis indicates that 92.4 % of RNA backbone 'suites' (sets of two consecutive sugar puckers with 5 connecting backbone torsions) fall into 54 rotameric classes, many of which are correlated with unique functions[12]. Non-rotameric suites are thus potential fitting errors. ERRASER-PHENIX reduced the number of such outliers in 22 of 24 cases and the average outlier rate from 19 % to 8 % (Table 1, Supplementary Table 4 and Fig. 2b). This result was particularly striking since the 54-rotamer classification was not used during the Rosetta modeling. In high-resolution cases, the ERRASER-fitted conformer typically agreed better with the electron density by visual inspection (Fig. 1b). For cases with medium-to-low resolution where the starting and remodeled conformer fit the density equally well visually, ERRASER-PHENIX gave substantially more rotameric conformers (Fig. 1c). As an additional test, we applied ERRASER during a recent RNA-puzzles blind trial[13] involving a protein-RNA complex. ERRASER-PHENIX changed a suite in the protein-binding kink-turn in starting RNA template (2YGH), from an outlier to the '2[' rotamer consistent with other kink-turn motifs[12] (Fig. 1d), which was indeed recovered in the subsequently released crystal structure (3V7E).

Fourth, RNA sugar rings typically exhibit either 2′-endo or 3′-endo conformations, but crystallographic assignments of these puckers can be ambiguous. While sugar pucker errors can be confidently identified using simple geometric criteria, finding alternative error-free solutions remains difficult[11]. ERRASER-PHENIX reduced the mean pucker error rate from 5 % to 0.2 %, and gave zero pucker errors in 19 cases (Table 1, Supplementary Table 4 and Fig. 2c). As an example with functional relevance, an adenosine in the active site of the group I ribozyme was fitted with different puckers in independent crystallographic models from bacteriophage Twort (A119 in 1Y0Q) and *Azoarcus* sp. BH72 (A127 in 3BO3). This discrepancy also led to different hydrogen bonding patterns between the adenosine's 2′-OH group and the guanosine (ΩG) substrate of the ribozyme (Fig. 1e). ERRASER-PHENIX improved agreement between the Twort and *Azoarcus* models throughout the active site and gave the same 2′-endo pucker conformation and hydrogen-bonding network (Fig. 1e), in agreement with recent double-mutant analyses of group I ribozyme[14].

In addition to correcting four classes of MolProbity-identified geometric problems, ERRASER-PHENIX improved other categories of errors. The ERRASER-PHENIX remodeling gave RNA base-pairing patterns with enhanced co-planarity and hydrogen-bonding geometry of interacting bases, as assessed by the automated base-pair assignment program MC-Annotate[15] and illustrated here by the a glycine riboswitch example (3P49, Fig. 1f). Furthermore, ERRASER-PHENIX led to remodeling of glycosidic bond torsions (syn vs. anti χ). In cases where higher resolution structures were available, the accuracy of these changes could be confirmed. Complete discussions are given in Supplementary Results and Supplementary Table 5-6.

In addition to the above improvements of geometric features, we also evaluated the fits of our models to the diffraction data using $R$ and $R_{free}$ factors. Avoiding increases in $R_{free}$, the correlation to set-aside diffraction data, is critical for preventing overfitting of the experimental data[16]. The ERRASER-PHENIX pipeline consistently decreased both $R$ and $R_{free}$, lowering $R_{free}$ in 22 out of 24 cases. The average $R$ dropped from 0.210 to 0.199 and average $R_{free}$ dropped from 0.255 to 0.243 (Table 1, Supplementary Table 7-8 and Fig. 2d). Other methods gave the same or worse average $R_{free}$. As a practical demonstration, we applied ERRASER-PHENIX to a newly solved structure of subdomain IIa from the hepatitis C virus internal ribosome entry site[17]. The ERRASER-PHENIX model gave fewer errors in all MolProbity criteria and lower $R$ and $R_{free}$, and was therefore deposited into PDB as the final structure (3TZR).

As a separate independent assessment, we compared the similarity of remodeled low-resolution structures to original PDB-deposited models of high-resolution structures with the same sequences. We reasoned that pairs of models with the same sequences should give similar local conformations, and the higher-resolution models could be used as working references. For all 13 such cases (Table 1, Supplementary Table 9-10), ERRASER-PHENIX remodeling gave low-resolution models with increased agreement in backbone torsions and sugar puckers to the deposited high-resolution models. In addition, we evaluated structures related by non-crystallographic symmetry or by internal homology and found that ERRASER improved their agreement in all tested cases (see Supplementary Results and Supplementary Table 11-12).

The quality improvement for lower-resolution models by ERRASER-PHENIX is further illustrated by comparison of the six datasets with worst diffraction resolution (3.20-3.69 Å) with five datasets at high-resolution (1.90-2.21 Å). For the low-resolution datasets, ERRASER-PHENIX improved the mean clashscore from 40.8 to 7.9, lower than the mean clashscore of 9.3 in the original high-resolution models. This value (7.9) is equal to the median clashscore for models solved at 1.8 Å in a recent whole PDB survey[3]. Similar reductions in outlier bond lengths and angles, outlier backbone rotamers, and anomalous sugar puckers are apparent (Supplementary Table 2-4).

For RNA crystallographic datasets across a wide range of resolutions and molecular size, ERRASER-PHENIX has led to consistent and substantial reduction of geometric errors, as assessed by independent validation tools and, in some cases, by independent functional evidence. The improved models give similar or better fits to set-aside diffraction data in all cases. For all geometric features, $R$ and $R_{free}$ values, the ERRASER-remodeled coordinates are significantly improved compared to starting PDB values ($P < 0.02$ by Wilcoxon signed-rank test; see Supplementary Table 13). Finally, comparison of remodeled low resolution and independent high resolution datasets indicates that this automated pipeline consistently increases the accuracy of RNA crystallographic models. We therefore expect this algorithm to mark an application of *ab initio* RNA 3D prediction that will be widely useful in experimental biology. ERRASER is available in the current Rosetta release (3.4) at http://www.rosettacommons.org, as an online application through ROSIE (Rosetta Online Server that Includes Everyone, http://rosie.rosettacommons.org), and as a part of the PHENIX package (http://www.phenix-online.org/).

**Acknowledgments**

We thank J. S. Richardson for suggesting this problem and for detailed evaluation of the results which we used to improve the program; C. L. Zirbel and N. B. Leontis for suggestions on base-pair validation; B. Stoner and D. Herschlag for discussions on group I ribozyme active site; T. Terwilliger and J. Headd for aids in integrating ERRASER into PHENIX; S. Lyskov for setting up the ERRASER protocol on the ROSIE Server; the Das lab for comments on the manuscript; and members of the Rosetta and the PHENIX communities for discussions and code sharing. Computations were performed on the BioX$^2$ cluster (NSF CNS-0619926) and XSEDE resources (NSF OCI-1053575). This work is supported by funding from NIH (R21 GM102716 to R.D. and R01 AI72012 to T.H.), a Burroughs-Wellcome Career Award at Scientific Interface (R.D.), Governmental Scholarship for Study Abroad of Taiwan and HHMI International Student Research Fellowship (F.C.C.), and the C. V. Starr Asia/Pacific Stanford Graduate Fellowship (P.S.).

**Author contributions**

F.C.C., P.S. and R.D. designed the research. F.C.C. implemented the methods and analyzed the results. P.S. provided code and assisted in data analysis. S.M.D and T.H. provided the starting model and diffraction data of the unreleased 3TZR structure and evaluated its refinement. F.C.C. and R.D. prepared the manuscript. All authors reviewed the manuscript.

## Competing financial interests

The authors declare no competing financial interests.

**Table 1**. Average values for the validation results of the benchmark set.

| | Outlier bond (%)[a] | Outlier angle (%)[a] | Clash score[b] | Outlier backbone rotamer (%)[c] | Potentially incorrect pucker (%)[d] | $R$ | $R_{free}$ | Nucleotide similarity (%)[e] | Pucker similarity (%)[e] |
|---|---|---|---|---|---|---|---|---|---|
| PDB | 0.53 | 1.18 | 18.03 | 18.8 | 5.0 | 0.210 | 0.256 | 64.9 | 91.5 |
| PHENIX | 0.01 | 0.03 | 10.79 | 15.2 | 2.4 | 0.199 | 0.244 | 71.7 | 96.4 |
| RNABC -PHENIX | 0.01 | 0 | 10.03 | 15.3 | 2.4 | 0.200 | 0.244 | 71.9 | 96.3 |
| RCrane -PHENIX | 0.003 | 0.12 | 10.12 | 10.3 | 1.0 | 0.207 | 0.252 | 74.1 | 95.8 |
| ERRASER -PHENIX | 0 | 0 | 7.04 | 7.9 | 0.2 | 0.199 | 0.244 | 80.5 | 97.0 |

[a] Bond lengths and angles that have a deviation > 4 s.d. compared to PHENIX ideal geometry[11].
[b] The number of serious clashes (atom-pairs that have steric overlaps ≥ 0.4 Å) per 1,000 atoms[11].
[c] Assigned using the definition from RNA Ontology Consortium[12].
[d] Determined using a geometric criterion based on the distance between the glycosidic bond vector (C1′–N1/9) and the following (3′) phosphate[11].
[e] Comparison of refined low-resolution models to independent high-resolution models (Supplementary Table 9). Nucleotides in which the differences between all torsion angles were smaller than 40° were denoted 'similar'. Nucleotides in which torsion angle δ agreed to within 20° were assigned 'similar' puckers.

**Figure 1**. Examples of geometric improvements by ERRASER-PHENIX. (**a**) Clash reduction in 3R4F. Red dots: unfavorable clashes. Left: PDB. Right: ERRASER-PHENIX. (**b**, **c**, **d**) Backbone conformation improvement on (**b**) nucleotides 62-64, chain A of 1U8D, (**c**) nucleotides 27-34, chain Q of 2OIU and (**d**) nucleotides 33-36, chain A of 2YGH. Rotamer assignments are shown at each suite. '!!' stands for outlier suites. Red: PDB. Blue: ERRASER-PHENIX. (**e**) Functionally relevant pucker correction on group I ribozyme models. Brown: 1Y0Q. Cyan: 3BO3. Left: PDB. Right: ERRASER-PHENIX. (**f**) Base-pair geometry improvement on nucleotides 1-6 and 66-71, chain A of 3P49. Left: PDB. Right: ERRASER-PHENIX.

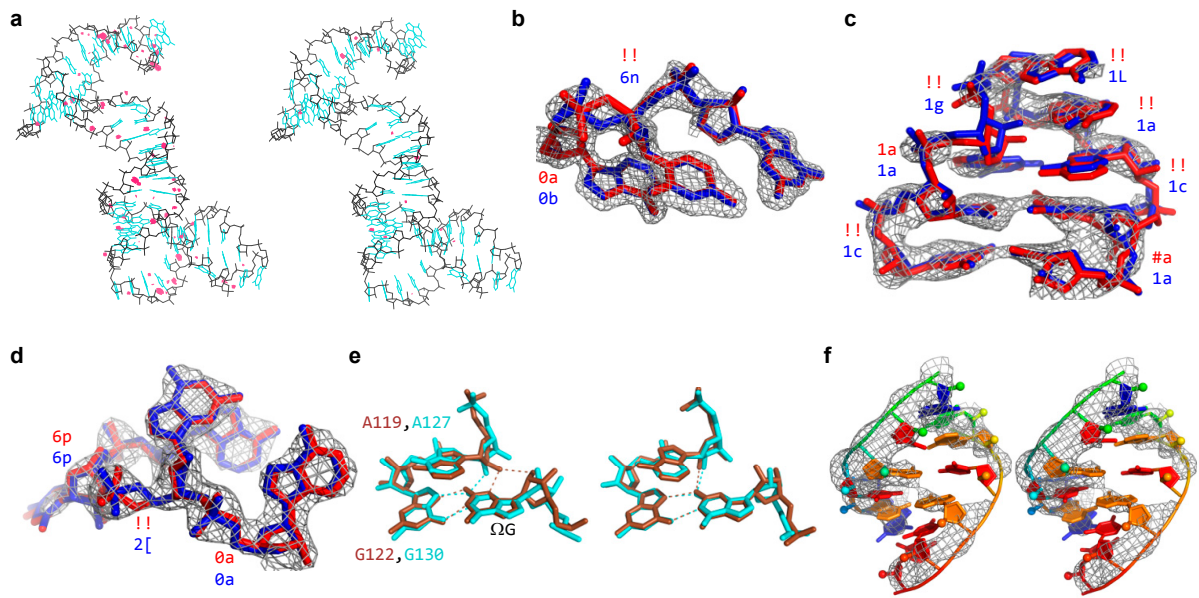

**Figure 2.** Improvements of the crystallographic models by ERRASER-PHENIX across the test cases. (**a**) clashscore, (**b**) frequencies of outlier backbone rotamers, (**c**) frequencies of outlier puckers and (**d**) $R_{free}$ factors (in percentage). The dashed lines give linear fits.

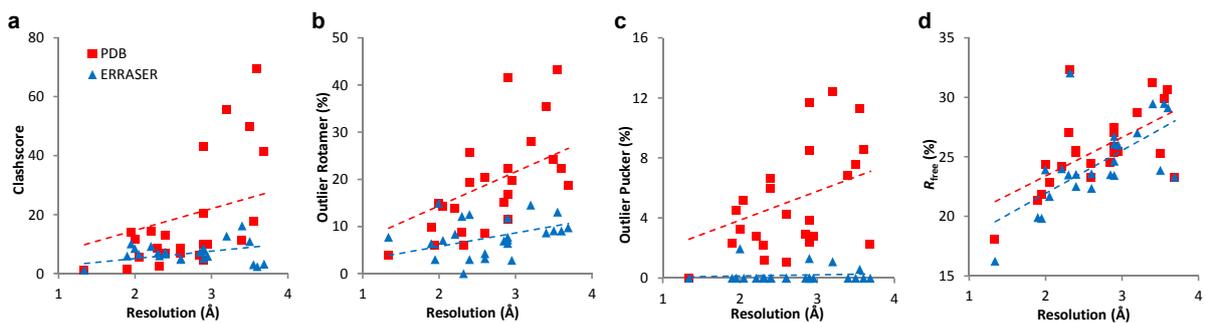

## Methods

*Overview of the ERRASER-PHENIX pipeline*

The ERRASER-PHENIX pipeline involves three major stages (Supplementary Fig. 1a). The starting model deposited in the PDB was first refined in PHENIX (v. dev-1034), with hydrogen atoms added. The refined model and electron-density map (setting aside the data for $R_{\text{free}}$ factor calculations; see below) were then passed into Rosetta (v. r50831) for a three-step real-space refinement. First, all torsion angles and all backbone bond lengths and bond angles were subjected to continuous minimization under the Rosetta high-resolution energy function supplemented with electron density correlation score. The Rosetta all-atom energy function models hydrogen bonding, Lennard-Jones packing, solvation, and torsional preferences, and has been successful in the modeling and design of RNA at near-atomic accuracy[7, 8]. The electron density score term is similar to the Rosetta electron density score recently pioneered for application to electron cryomicroscopy and molecular replacement[9, 10]. Second, bond length, bond angle, pucker and suite outliers were identified using phenix.rna_validate. In addition, we also included nucleotides that shifted substantially during the initial Rosetta minimization (evaluated by nucleotide-wise RMSD before and after minimization). These outlier and high-RMSD nucleotides were rebuilt by single-nucleotide StepWise Assembly (SWA) in a one-by-one fashion, where all of a nucleotide's atoms and the atoms up to the previous and next sugar were sampled by an exhaustive grid search of all torsions and a kinematic loop closure at sub-Angstrom resolution (Supplementary Fig. 1b)[8, 18]. If SWA found a lower-energy alternative structure of the rebuilt nucleotide, this new conformation was accepted. Third, the new model was minimized again in Rosetta. The rebuilding-minimization cycle was iterated three times to obtain the final ERRASER model. This model was again refined in PHENIX against diffraction data to obtain the final ERRASER-PHENIX model. All the ERRASER-PHENIX remodeled structures discussed in this research are available as Supplementary Data.

*The new Rosetta module, ERRASER*

The ERRASER protocol consisted of three steps: an initial whole structure minimization, followed by single nucleotide rebuilding, and finally another whole structure minimization. Before passing the models into ERRASER, the PHENIX-generated pdb files were converted to the Rosetta format. Protein components, ligands and modified nucleotides were removed from the model, because current enumerative Rosetta modeling only handles standard RNA nucleotides. To avoid anomalies in refitting, we held fixed the positions of the nucleotides that were bonded or in van der Waals contact with these removed atoms during the next ERRASER step. In 2OIU, a cyclic RNA structure, we also held fixed the first and the last nucleotides in the RNA chain to prevent the bonds from breaking during ERRASER. For structures that have notable interaction through crystal contacts, we manually included the interacting atoms into the ERRASER starting models.

Throughout the ERRASER refinement, an electron density score (unbiased by excluding set-aside $R_{free}$ reflections during map creation, see below) was added to the energy function to ensure that the rebuilt structural models retained a reasonable fit to the experimental data. The electron density scoring in our method is slightly different from the one published recently [9, 10]. Instead of calculating the density profile of the model every time we rescored the model, we pre-calculated the correlation between the density of a single atom and the experimental density in a fine grid. The score was defined as the negative of the sum of the atomic numbers of all the heavy atoms in the model times this rapidly computed real-space correlation coefficient. This new density scoring term, named `elec_dens_atomwise`, was an order of magnitude faster than the one in the previous Rosetta release, thus reducing the total computational time of our method substantially. To accommodate the change of our energy function caused by the electron density energy constraint, we also modified the weights in the original scoring function. The scoring weights file is included in the Rosetta release named `rna_hires_elec_dens.wts`.

In addition, we used a new RNA torsional potential for this study. This new potential was obtained by fitting to the logarithm of the histogram of RNA torsions derived from the RNA11 dataset (http://kinemage.biochem.duke.edu/databases/rnadb.php ). The RNA11 dataset contains 24,842 RNA suites and 311 different pdb entries, which is much richer and more diverse than the 50S ribosomal subunit model (1JJ2, 2,875 suites) used in deriving the original potential[7]. This new potential can be turned on by including the tag "`-score:rna_torsion_potential RNA11_based_new`" in the Rosetta command line (see Supplementary Notes)

During the whole structure minimization, we constrained the phosphate atoms in the RNA to their starting position; this is especially important for low-resolution models where the phosphate positions were not accurately defined by electron density. Errors in phosphate positions were corrected during the latter rebuilding step. We also found that when the molecule was too large, Rosetta was unable to minimize the entire molecule due to slow scoring. Therefore for any molecule larger than 150 nucleotides, we decomposed the RNA into smaller segments with an automated script `rna_decompose.py`, and minimized each of them sequentially. To retain all interactions, we also included the nucleotides within 5 Å radius of the segment being minimized as fixed nucleotides during the minimization.

After the whole structure minimization, we used phenix.rna_validate to analyze the obtained models. All nucleotides assigned to have outlier bond lengths, bond angles, puckers and/or potentially erroneous backbone rotamers (outliers or regular rotamers with suiteness < 0.1; suiteness is a quality measurement for rotamer assignments[12]) were identified as problematic and were rebuilt in subsequent Rosetta single nucleotide rebuilding. Furthermore, because the single nucleotide rebuilding region in Rosetta did not match the definition of a "suite", we rebuilt both the selected nucleotide and the nucleotide preceding it to cover the whole suite for rotameric outliers.

In addition to rebuilding outlier nucleotides, we also computed the nucleotide-wise RMSD between the models before and after minimization. The nucleotides with RMSD larger than 0.05 times the diffraction resolution and within the 20 % of the largest RMSD nucleotide were also selected for rebuilding. We reasoned that because these nucleotides moved substantially after Rosetta minimization, their starting conformations were not favorable in terms of Rosetta energy function and were potentially erroneous.

The single nucleotide rebuilding step used in our method was based on a modified SWA algorithm in which the RNA chain was closed using triaxial kinematic loop closure[18]. For nucleotides at chain termini, the original SWA sampling was used since no chain closure was required. For rebuilding nucleotides inside the RNA chain, we first created a chain break between O3' and P in the lower suite of the rebuilding nucleotide. Then we sampled all possible torsion angles for $\varepsilon_i, \zeta_i, \alpha_i, \alpha_{i+1}$ in 20° steps, and the two most common conformation of the sugar pucker, 2′-endo and 3′-endo. For each sampled conformation, analytical loop closure was applied to close the chain and determine the remaining 6 torsions ($\beta_i, \gamma_i, \varepsilon_i, \zeta_i, \beta_{i+1}, \gamma_{i+1}$) which form three pairs of pivot-sharing torsions. The glycosidic torsion $\chi_i$ and the 2′-OH torsion $\chi_i^{2'-OH}$ were sampled after chain closure, and the generated models were further minimized in Rosetta. During the rebuilding, we applied a modest constraint to the glycosidic torsion so that it is more stable near the starting conformation, therefore only the base-orientation changes that gave substantial Rosetta energy bonuses were accepted as the final conformations. To reduce the computational expense, we only searched conformations that were within 3.0 Å RMSD with respect to the starting models.

After the conformational search, 100 lowest energy models were kept and further minimized under the constraint of the Rosetta `linear_chainbreak` and `chainbreak` energy term to ensure that the chain break was closed properly in the final model. Finally the best scored model was outputted as the new model for the RNA. If no new low-energy model could be found, then the program kept the starting model of that nucleotide. In the rebuilding process, the candidate nucleotides were rebuilt sequentially from the 5′-end to 3′-end of the RNA sequence. In order to speed up the Rosetta rebuilding process, the nucleotide being rebuilt was cut out from the whole structural model together with all nucleotides within 5 Å radius, rebuilt using SWA, and pasted back to the model.

After all the problematic nucleotides were rebuilt, we minimized the whole model again to further reduce any bond length or angle errors that might have occurred in the rebuilding process, and to improve the overall energy of the model. In this study, the rebuilding-minimization cycle was iterated three times, although single iterations gave nearly equivalent results (not shown). The coordinates of the RNA atoms in the PHENIX model were then substituted by the new coordinates in the Rosetta-rebuilt model to give the final ERRASER output.

The three ERRASER steps discussed above were wrapped into a python script `erraser.py` and can be performed automatically. The user needs to input a starting pdb file, a ccp4 map file, the resolution of the map and a list of any nucleotides that should be held fixed during refinement due to their interaction with removed atoms.

A sample ERRASER command line used for the refinement of 3IWN is shown below:

```
erraser.py -pdb 3IWN.pdb -map 3IWN.ccp4 -map_reso 3.2 -fixed_res
A37 A58-67 B137 B158-167
```

Here `3IWN.pdb` is the name of PHENIX refined model, `3IWN.ccp4` is the name of ccp4 density map file, `-map_reso` tag gives the resolution of the density map, and `-fixed_res` specifies the nucleotides that should remain untouched. "A37" means the 37[th] nucleotide of chain A in the pdb file.

Examples of the automatically generated Rosetta command lines by the python script are given in Supplementary Notes.

*PHENIX refinement*

PHENIX[19] (v. dev-1034) was used for all the refinements performed in this study. We first prepared the starting models downloaded from the PDB for refinement using phenix.ready_set. This step added missing hydrogen atoms into the models and set up constraint files including ligand constraints and metal coordination constraints. For ligands A23, 1PE and CCC, we substituted the PHENIX-generated ligand constraints with constraint files from the CCP4 monomer library to achieve better geometry. Furthermore, phenix.ready_set did not automatically create bond length and bond angle constraints at the linkage between some modified nucleotides (GDP and GTP) and standard nucleotides, or between the first and the last nucleotide of a cyclic RNA. In such cases these constraints were added manually. Finally, for pdb files with TLS (Translation, Libration, Screw)[20] refinement records, the TLS group information was manually extracted from the pdb files and saved in a separate file for further use in PHENIX.

After all the files for the refinement were ready, a four-step PHENIX refinement was performed. In the first step, because PHENIX does not load in TLS records in the pdb files, we performed a one-cycle TLS refinement to recover the TLS information. Second, the models were refined by phenix.refine for three cycles. At this step, ADP (Atomic Displacement Parameters) weight (wxu_scale) was optimized by PHENIX using a grid search, and other parameters were manually determined based on the criteria described below. For higher resolution structures a higher wxc_scale (scale for X-ray vs. Sterochemistry weight) was found to be appropriate. Based on initial tests (on PDB cases [1Q9A](1Q9A) and [2HOP](2HOP), which were not included in this paper's benchmark since they were used to set parameters), we used the following criteria: wxc_scale = 0.5 for

Resolution < 2.3 Å, wxc_scale = 0.1 for 2.3 Å ≤ Resolution < 3 Å, wxc_scale = 0.05 for 3 Å ≤ Resolution ≤ 3.6 Å, and wxc_scale = 0.03 for Resolution > 3.6 Å. The ordered_solvent option (automatic water updating) was turned on for all structures. Empirically, we found that the real-space refinement strategy in PHENIX only gave equal or worse $R$ factor, so it was turned off throughout all the refinement steps in this study. TLS refinement was turned on only for structures with TLS record in the deposited PDB files. Third, the models were further refined in phenix.refine for nine cycles using the same parameter set. Fourth, the models were further refined in phenix.refine for three cycles, with all target weights (wxc_scale and wxu_scale) optimized during the run. Other parameters stayed the same as in the first refinement round. Finally, we compared the models by the three different refinement steps and selected the one with best fit to the diffraction data as the final model. For 3OTO, the multi-step PHENIX refinement clearly distorted the starting model and gave worse geometries, so in this case we used the results obtained after the first refinement step. For 3P49, we supplied 1URN as a reference model to improve the protein part of the structure during refinement[21].

After the initial refinement, the electron density map was generated from the experimental diffraction data and the PHENIX refined structural model for further ERRASER improvement. We used phenix.maps to create $2mF_{obs}$-$DF_{calc}$ maps in ccp4 format, and diffraction data used for $R_{free}$ validation were excluded for the map generation to avoid directly fitting to the $R_{free}$ test set during the ERRASER refinement. To avoid Fourier truncation errors due to the missing data, we filled any excluded or missing $F_{obs}$ values with $F_{calc}$ values during the map calculation. The averaged kicked map approach was also used to reduce the noise and model bias of the maps[22]. An example for the input file used in map creation is given below.

The final PHENIX refinement, after the ERRASER steps, was similar to the starting refinement described above, with small variations. First, there was no need for an initial TLS refinement since the pdb files already had this information at this stage. Second, we ran phenix.ready_set again on the ERRASER model to generate metal coordination constraints for refinements, in case the new model presented different metal coordination patterns than the starting one. The models were then refined using PHENIX in the same multi-step fashion, with the same parameter sets.

Examples of the PHENIX command lines used in this work are given in Supplementary Notes.

*Refinement of 3TZR, a new structure of subdomain IIa from the hepatitis C virus IRES domain*
The refinement of the 3TZR model currently deposited in the PDB was performed at an earlier stage of this work using an earlier PHENIX version (v1.7.1-743). The initial coordinates for 3TZR were already well-refined in PHENIX, and we therefore maintained the settings from that initial stage. In particular, during the PHENIX refinement, hydrogen atoms were not added to the model, and wxc_scale was set to 0.5. The final PHENIX refinement was performed using the same setting as the initial one.

*R and $R_{free}$ calculation*
For consistency, $R$ and $R_{free}$ values of all the models were calculated using phenix.model_vs_data[23]. For the starting models, the PHENIX-calculated $R$ and $R_{free}$ were generally similar to the values shown in the PDB header; both are reported in Supplementary

Table 7-8. In the main text, we have reported PHENIX-calculated $R$ and $R_{free}$ to permit comparisons across the refinement benchmark.

*Similarity analysis test*

The similarities of the local geometries between similar structural models (Supplementary Table 9-12) were evaluated as follows. If differences between the torsion angles ($\alpha, \beta, \gamma, \delta, \varepsilon, \zeta, \chi$) of each nucleotide pair were all smaller than 40°, the pair was counted as a similar nucleotide pair. If the difference of the $\delta$ angles of a nucleotide pair was smaller than 20°, the pair was assigned as having similar sugar pucker. Finally, RMSDs of all the torsion angles (in degrees) between the model pairs were calculated as an indicator of the model similarity in the torsional space.

*Other tools*

RNABC[4] (v1.11) and RCrane.CLI[5] (v1.01) were combined with PHENIX in the same manner as the ERRASER-PHENIX pipeline, by substituting the ERRASER stage with RNABC and RCrane, respectively. Since RNABC rebuilt only one nucleotide per run, a python script was used to achieve automatic rebuilding of all nucleotides. The MolProbity[11] analysis was performed using command line tools phenix.clashscore and phenix.rna_validate in the PHENIX package. MC-Annotate[15] (v1.6.2) was used to assign base-pairs in starting and refined models. All molecular images in this work were prepared using PyMol, except Figure 1a, which used MolProbity[11] and KiNG (Kinemage, Next Generation)[24].

*Supplementary Information*

# Correcting pervasive errors in RNA crystallography through enumerative structure prediction

Fang-Chieh Chou, Parin Sripakdeevong, Sergey M. Dibrov, Thomas Hermann & Rhiju Das[*]

| Supplementary Item | Title |
|---|---|
| Supplementary Figure 1 | ERRASER-PHENIX algorithm. |
| Supplementary Figure 2 | Base orientation improvements. |
| Supplementary Table 1 | Benchmark test set of 24 structural models, sorted by resolution. |
| Supplementary Table 2 | Outlier bond lengths and angles of the benchmark assessed by phenix.rna_validate. |
| Supplementary Table 3 | Clashscore of the benchmark assessed by phenix.clashscore. |
| Supplementary Table 4 | Outlier backbone rotamers and potentially incorrect sugar puckers of the benchmark assessed by phenix.rna_validate. |
| Supplementary Table 5 | Number of base-pairs identified by MC-Annotate. |
| Supplementary Table 6 | Summary of notable changes of base orientations ($\chi$ angle). |
| Supplementary Table 7 | $R$ factors of the benchmark. |
| Supplementary Table 8 | $R_{free}$ factors of the benchmark. |
| Supplementary Table 9 | Similarity analysis between high-resolution and low-resolution models. |
| Supplementary Table 10 | Torsional RMSDs (in degrees) between high-resolution and low-resolution models. |
| Supplementary Table 11 | Similarity analysis for model pairs of the same or similar sequences. |
| Supplementary Table 12 | Torsional RMSDs (in degrees) for model pairs of the same or similar sequences. |
| Supplementary Table 13 | P-values of Wilcoxon signed-rank test between each method and the starting PDB dataset for all geometric features, $R$ and $R_{free}$. |
| Supplementary Notes | Example Rosetta command lines used in ERRASER-PHENIX. |
| Supplementary Results | |

**Supplementary Figure 1**. ERRASER-PHENIX algorithm. (**a**) Flow chart of the whole pipeline, the ERRASER steps are enclosed in red. (**b**) Enumerative RNA modeling in Rosetta steps. The blue torsions are explicitly sampled by enumeration and the green torsions are determined by automatic loop closure.

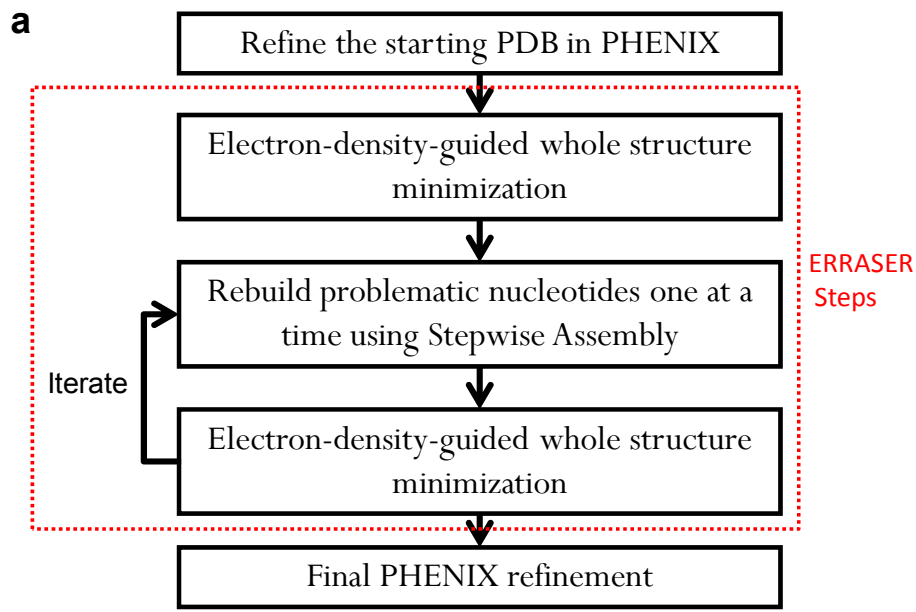

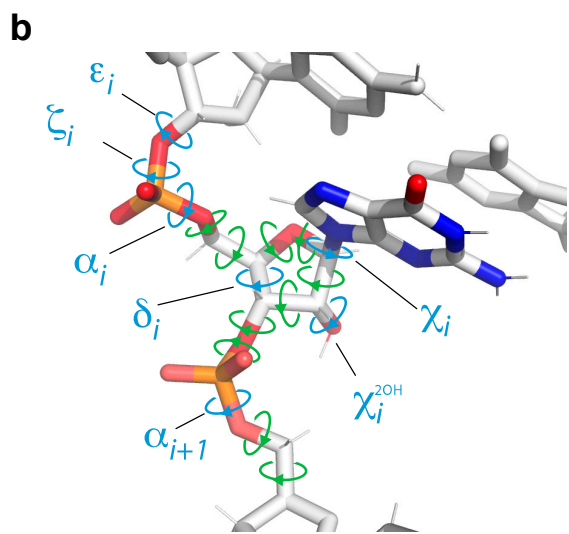

**Supplementary Figure 2**. Base orientation improvements. (**a**) Probability distribution of the χ angle in RNA09 (http://kinemage.biochem.duke.edu/databases/rnadb.php). The vertical lines show the range of syn vs. anti. Red: Purine; Blue: Pyrimidine; Dotted lines: 20X zoom-in of the distributions. (**b**) Base orientation changes in 2CKY, chain A, residue U35. Red: PDB model; Blue: ERRASER-PHENIX model; Brown: Reference model. (**c, d**) Base orientation changes in the ribosomal subunit 3OTO for residue (**c**) G266 and (**d**) U365. Red: PDB model; Blue: ERRASER-PHENIX model; Brown: Reference model (2VQE). Upper panel: Plot with electron density from 3OTO. Lower panel: Plot with electron density from 2VQE; the ERRASER model is aligned to the 2VQE model.

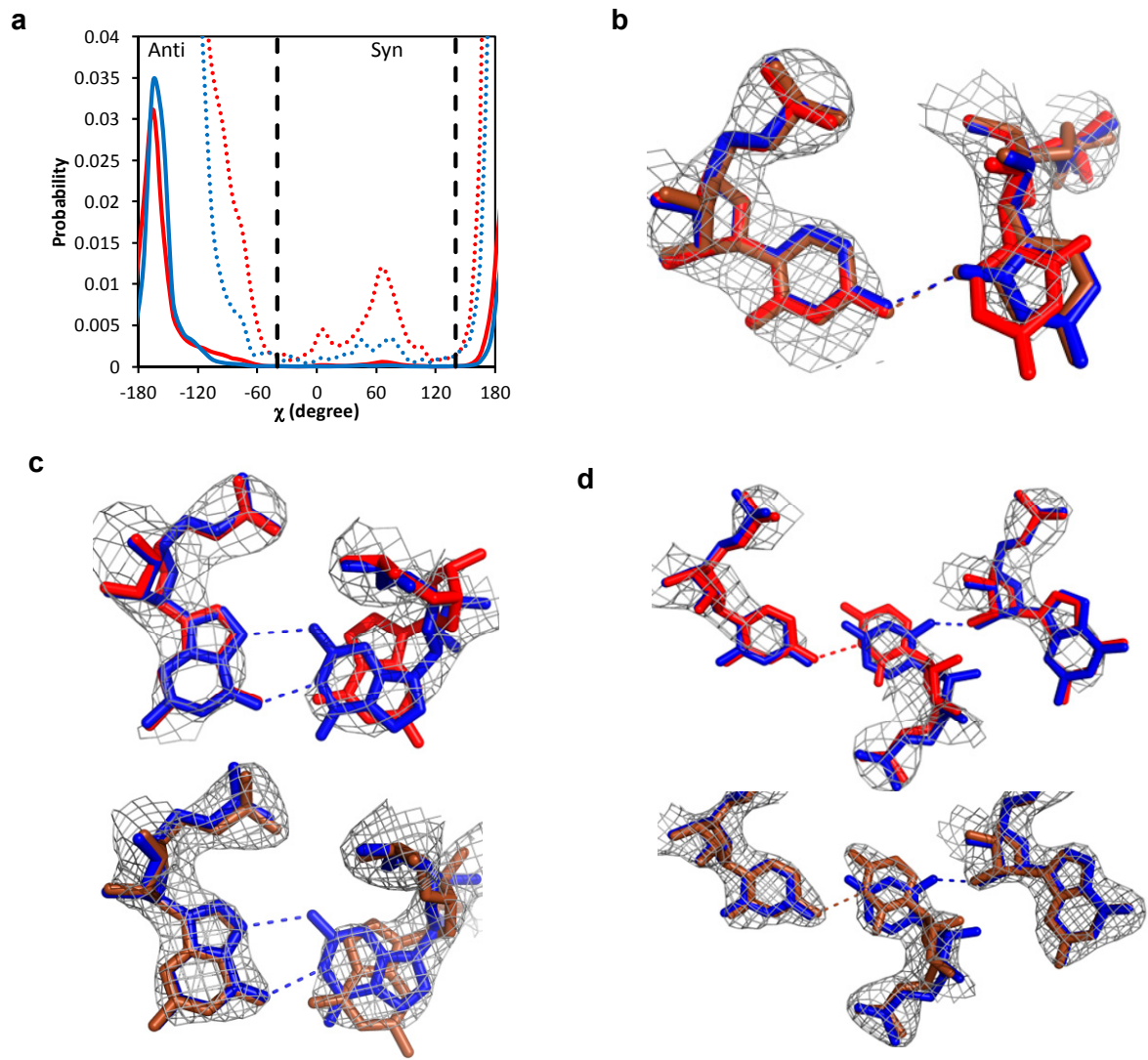

**Supplementary Table 1.** Benchmark test set of 24 structural models, sorted by resolution.

| PDB ID | Name | RNA length[a] | Resolution (Å) |
|---|---|---|---|
| 2A43 | RNA Pseudoknot | 26 | 1.34 |
| 3DIL | Lysine Riboswitch | 174 | 1.9 |
| 1U8D | Guanine Riboswitch | 68 | 1.95 |
| 3D2V | Eukaryotic TPP Riboswitch | 77 × 2 | 2 |
| 2GDI | E. coli. TPP Riboswitch | 80 × 2 | 2.05 |
| 3TZR | HCV IRES Subdomain IIa | 36 | 2.21 |
| 3MXH | c-di-GMP Riboswitch | 92 | 2.3 |
| 2PN4 | HCV IRES Subdomain IIa | 44 × 2 | 2.32 |
| 2QUS | Hammerhead Ribozyme | 69 × 2 | 2.4 |
| 1Y27 | Guanine Riboswitch | 68 | 2.4 |
| 2OIU | L1 Ribozyme Ligase | 71 × 2 | 2.6 |
| 2YGH | SAM-I Riboswitch | 95 | 2.6 |
| 3DIZ | Lysine Riboswitch | 174 | 2.85 |
| 3E5E | SAM-III Riboswitch | 53 | 2.9 |
| 2GIS | SAM-I Riboswitch | 94 | 2.9 |
| 2CKY | Eukaryotic TPP Riboswitch | 77 × 2 | 2.9 |
| 2PN3 | HCV IRES Subdomain IIa | 44 | 2.9 |
| 3F2Q | FMN Riboswitch | 112 | 2.95 |
| 3IWN | c-di-GMP Riboswitch | 93 × 2 | 3.2 |
| 3BO3 | *Azoarcus* Group I Ribozyme | 212 | 3.4 |
| 3R4F | Prohead RNA | 66 | 3.5 |
| 3P49 | Glycine Riboswitch | 169 | 3.55 |
| 1Y0Q | Twort Group I Ribozyme | 233 | 3.6 |
| 3OTO | 30S Ribosomal Subunit | 1,522 | 3.69 |

[a] The length of RNA component of the molecule. "× 2" indicates 2 copies in the asymmetric unit.

**Supplementary Table 2.** Outlier bond lengths and angles of the benchmark assessed by phenix.rna_validate.

| PDB ID | Outlier bonds(%) | | | | | Outlier angles(%) | | | | |
|---|---|---|---|---|---|---|---|---|---|---|
| | PDB | PHENIX | RNABC-PHENIX | RCrane-PHENIX | ERRASER-PHENIX | PDB | PHENIX | RNABC-PHENIX | RCrane-PHENIX | ERRASER-PHENIX |
| 2A43 | 0 | 0 | 0 | 0 | 0 | 0 | 0 | 0 | 0 | 0 |
| 3DIL | 0 | 0 | 0 | 0 | 0 | 0 | 0 | 0 | 0 | 0 |
| 1U8D | 0 | 0 | 0 | 0 | 0 | 2.99 | 0 | 0 | 0 | 0 |
| 3D2V | 0 | 0 | 0 | 0 | 0 | 0 | 0 | 0 | 0 | 0 |
| 2GDI | 0 | 0 | 0 | 0 | 0 | 0 | 0 | 0 | 0 | 0 |
| 3TZR | 0 | 0 | 0 | 0 | 0 | 0 | 0 | 0 | 2.78 | 0 |
| 3MXH | 0 | 0 | 0 | 0 | 0 | 0 | 0 | 0 | 0 | 0 |
| 2PN4 | 2.38 | 0 | 0 | 0 | 0 | 2.38 | 0 | 0 | 0 | 0 |
| 2QUS | 1.47 | 0 | 0 | 0 | 0 | 3.68 | 0.74 | 0 | 0 | 0 |
| 1Y27 | 0 | 0 | 0 | 0 | 0 | 4.48 | 0 | 0 | 0 | 0 |
| 2OIU | 2.11 | 0 | 0 | 0 | 0 | 0.7 | 0 | 0 | 0 | 0 |
| 2YGH | 3.19 | 0 | 0 | 0 | 0 | 2.13 | 0 | 0 | 0 | 0 |
| 3DIZ | 0 | 0 | 0 | 0 | 0 | 0 | 0 | 0 | 0 | 0 |
| 3E5E | 1.92 | 0 | 0 | 0 | 0 | 0 | 0 | 0 | 0 | 0 |
| 2GIS | 0 | 0 | 0 | 0 | 0 | 6.38 | 0 | 0 | 0 | 0 |
| 2CKY | 1.3 | 0 | 0 | 0 | 0 | 0.65 | 0 | 0 | 0 | 0 |
| 2PN3 | 0 | 0 | 0 | 0 | 0 | 0 | 0 | 0 | 0 | 0 |
| 3F2Q | 0 | 0 | 0 | 0 | 0 | 0 | 0 | 0 | 0 | 0 |
| 3IWN | 0 | 0 | 0 | 0 | 0 | 4.3 | 0 | 0 | 0 | 0 |
| 3BO3 | 0.45 | 0 | 0 | 0 | 0 | 0 | 0 | 0 | 0 | 0 |
| 3R4F | 0 | 0 | 0 | 0 | 0 | 0 | 0 | 0 | 0 | 0 |
| 3P49 | 0 | 0 | 0 | 0 | 0 | 0.59 | 0 | 0 | 0 | 0 |
| 1Y0Q | 0 | 0 | 0 | 0 | 0 | 0 | 0 | 0 | 0 | 0 |
| 3OTO | 0 | 0.2 | 0.27 | 0.07 | 0 | 0 | 0.07 | 0 | 0.13 | 0 |
| Average | 0.53 | 0.01 | 0.01 | 0.003 | 0 | 1.18 | 0.03 | 0 | 0.12 | 0 |
| Avg. high-res[a] | 0 | 0 | 0 | 0 | 0 | 0.60 | 0 | 0 | 0.56 | 0 |
| Avg. low-res[b] | 0.08 | 0.03 | 0.05 | 0.01 | 0 | 0.82 | 0.01 | 0 | 0.02 | 0 |
| Equal to or better than PDB | | 23 / 24 | 23 / 24 | 23 / 24 | 24 / 24 | | 23 / 24 | 24 / 24 | 22 / 24 | 24 / 24 |

Bond lengths and angles that have a deviation > 4 σ compared to PHENIX ideal geometry are counted as outliers.
[a] Average value for the five high resolution models 3DIL, 1U8D, 3D2V, 2GDI and 3TZR. Ultra-high resolution dataset 2A43 was excluded.
[b] Average value for the six lowest resolution models 3IWN, 3BO3, 3R4F, 3P49, 1Y0Q and 3OTO.

**Supplementary Table 3.** Clashscore of the benchmark assessed by phenix.clashscore.

| PDB ID | PDB | PHENIX | RNABC-PHENIX | RCrane-PHENIX | ERRASER-PHENIX |
|---|---|---|---|---|---|
| 2A43 | 1.19 | 1.19 | 2.38 | 2.38 | 1.19 |
| 3DIL | 1.40 | 6.29 | 5.94 | 5.94 | 5.94 |
| 1U8D | 14.02 | 16.22 | 13.59 | 12.71 | 10.08 |
| 3D2V | 11.43 | 8.67 | 8.47 | 10.84 | 8.47 |
| 2GDI | 5.52 | 8.41 | 8.98 | 8.60 | 6.50 |
| 3TZR | 14.24 | 13.40 | 14.24 | 14.24 | 9.21 |
| 3MXH | 8.48 | 5.58 | 6.69 | 8.93 | 6.92 |
| 2PN4 | 2.48 | 5.31 | 8.85 | 9.21 | 6.02 |
| 2QUS | 12.97 | 13.43 | 10.52 | 12.75 | 6.71 |
| 1Y27 | 6.82 | 10.00 | 6.36 | 5.91 | 7.27 |
| 2OIU | 8.48 | 7.61 | 7.61 | 9.13 | 5.00 |
| 2YGH | 6.78 | 4.53 | 7.11 | 5.16 | 4.85 |
| 3DIZ | 6.17 | 7.40 | 7.05 | 12.34 | 7.23 |
| 3E5E | 4.51 | 4.53 | 2.83 | 9.07 | 5.09 |
| 2GIS | 43.14 | 23.39 | 22.75 | 9.29 | 8.01 |
| 2CKY | 20.47 | 12.99 | 11.22 | 10.83 | 7.68 |
| 2PN3 | 9.94 | 7.81 | 7.10 | 7.81 | 8.52 |
| 3F2Q | 9.74 | 7.80 | 8.91 | 6.40 | 5.85 |
| 3IWN | 55.65 | 23.31 | 19.63 | 13.94 | 12.71 |
| 3BO3 | 11.05 | 13.35 | 15.61 | 12.56 | 16.23 |
| 3R4F | 49.62 | 16.07 | 15.12 | 16.54 | 10.85 |
| 3P49 | 17.56 | 5.43 | 6.57 | 7.43 | 3.00 |
| 1Y0Q | 69.53 | 20.44 | 7.96 | 16.19 | 2.39 |
| 3OTO | 41.42 | 15.75 | 15.30 | 14.69 | 3.17 |
| Average | 18.03 | 10.79 | 10.03 | 10.12 | 7.04 |
| Avg. high-res[a] | 9.32 | 10.60 | 10.24 | 10.47 | 8.04 |
| Avg. low-res[b] | 40.81 | 15.73 | 13.37 | 13.56 | 8.06 |
| Equal to or better than PDB | | 15 / 24 | 17 / 24 | 15 / 24 | 17 / 24 |

Clashscore is the number of serious clashes (atom-pairs that have steric overlaps ≥ 0.4 Å) per 1,000 atoms.
[a] Average value for the five high resolution models 3DIL, 1U8D, 3D2V, 2GDI and 3TZR. Ultra-high resolution dataset 2A43 was excluded.
[b] Average value for the six lowest resolution models 3IWN, 3BO3, 3R4F, 3P49, 1Y0Q and 3OTO.

**Supplementary Table 4.** Outlier backbone rotamers and potentially incorrect sugar puckers of the benchmark assessed by phenix.rna_validate.

| PDB ID | Outlier backbone rotamers | | | | | Potentially incorrect puckers | | | | |
|---|---|---|---|---|---|---|---|---|---|---|
| | PDB | PHENIX | RNABC-PHENIX | RCrane-PHENIX | ERRASER-PHENIX | PDB | PHENIX | RNABC-PHENIX | RCrane-PHENIX | ERRASER-PHENIX |
| 2A43 | 1 | 1 | 1 | 3 | 2 | 0 | 0 | 0 | 0 | 0 |
| 3DIL | 17 | 13 | 14 | 11 | 11 | 4 | 2 | 1 | 0 | 0 |
| 1U8D | 4 | 5 | 7 | 3 | 2 | 3 | 2 | 2 | 0 | 0 |
| 3D2V | 23 | 25 | 26 | 18 | 23 | 5 | 2 | 2 | 2 | 3 |
| 2GDI | 22 | 17 | 16 | 14 | 11 | 8 | 2 | 2 | 3 | 0 |
| 3TZR | 5 | 5 | 5 | 3 | 3 | 1 | 1 | 1 | 0 | 0 |
| 3MXH | 8 | 9 | 10 | 11 | 11 | 2 | 1 | 1 | 0 | 0 |
| 2PN4 | 5 | 2 | 1 | 1 | 0 | 1 | 0 | 0 | 0 | 0 |
| 2QUS | 35 | 26 | 24 | 18 | 17 | 9 | 6 | 6 | 2 | 0 |
| 1Y27 | 13 | 10 | 8 | 2 | 2 | 4 | 2 | 2 | 0 | 0 |
| 2OIU | 29 | 17 | 17 | 8 | 6 | 6 | 0 | 0 | 0 | 0 |
| 2YGH | 8 | 7 | 8 | 5 | 3 | 1 | 0 | 0 | 0 | 0 |
| 3DIZ | 26 | 19 | 17 | 11 | 12 | 5 | 1 | 1 | 0 | 0 |
| 3E5E | 6 | 6 | 5 | 4 | 4 | 2 | 1 | 1 | 0 | 0 |
| 2GIS | 21 | 16 | 20 | 14 | 6 | 8 | 7 | 7 | 1 | 0 |
| 2CKY | 64 | 44 | 37 | 21 | 18 | 18 | 5 | 5 | 2 | 2 |
| 2PN3 | 7 | 2 | 2 | 2 | 3 | 1 | 0 | 0 | 0 | 0 |
| 3F2Q | 21 | 15 | 15 | 9 | 3 | 3 | 2 | 2 | 2 | 0 |
| 3IWN | 52 | 42 | 42 | 29 | 27 | 23 | 12 | 11 | 4 | 2 |
| 3BO3 | 78 | 70 | 63 | 22 | 19 | 15 | 6 | 6 | 1 | 0 |
| 3R4F | 16 | 13 | 17 | 10 | 6 | 5 | 1 | 1 | 1 | 0 |
| 3P49 | 73 | 57 | 60 | 31 | 22 | 19 | 11 | 11 | 8 | 1 |
| 1Y0Q | 52 | 59 | 58 | 55 | 21 | 20 | 14 | 11 | 8 | 0 |
| 3OTO | 279 | 301 | 313 | 253 | 145 | 34 | 35 | 37 | 27 | 0 |
| Average (%)[a] | 18.8 | 15.2 | 15.3 | 10.3 | 7.9 | 5.0 | 2.4 | 2.4 | 1.0 | 0.2 |
| Avg. high-res[b] | 11.7 | 11.2 | 11.9 | 8.0 | 7.9 | 3.6 | 1.9 | 1.8 | 0.6 | 0.4 |
| Avg. low-res[c] | 28.6 | 25.6 | 26.4 | 16.6 | 10.7 | 8.1 | 4.3 | 4.0 | 2.3 | 0.3 |
| Equal to or better than PDB | | 19 / 24 | 18 / 24 | 21 / 24 | 22 / 24 | | 23 / 24 | 23 / 24 | 24 / 24 | 24 / 24 |

The backbone rotamer assignment is an effort of RNA Ontology Consortium[1]. Sugar pucker errors are determined using a geometric criterion based on the distance between the glycosidic bond vector (C1′–N1/9) and the following (3′) phosphate[2].

[a] Average error rate as percentage. Obtained by dividing the number of outliers with the total number of nucleotides.
[b] Average value for the five high resolution models 3DIL, 1U8D, 3D2V, 2GDI and 3TZR. Ultra-high resolution dataset 2A43 was excluded. The values are normalized by the numbers of nucleotides in the models.
[c] Average value (normalized) for the six lowest resolution models 3IWN, 3BO3, 3R4F, 3P49, 1Y0Q and 3OTO.

**Supplementary Table 5.** Number of base-pairs identified by MC-Annotate.

| PDB ID | PDB | PHENIX | RNABC-PHENIX | RCrane-PHENIX | ERRASER-PHENIX |
|---|---|---|---|---|---|
| 2A43 | 12 | 12 | 12 | 12 | 12 |
| 3DIL | 88 | 89 | 89 | 90 | 91 |
| 1U8D | 31 | 32 | 32 | 32 | 32 |
| 3D2V | 69 | 68 | 68 | 69 | 68 |
| 2GDI | 70 | 70 | 70 | 71 | 70 |
| 3TZR | 15 | 15 | 15 | 15 | 15 |
| 3MXH | 38 | 40 | 40 | 38 | 40 |
| 2PN4 | 32 | 32 | 32 | 32 | 32 |
| 2QUS | 52 | 52 | 51 | 52 | 52 |
| 1Y27 | 36 | 34 | 35 | 34 | 36 |
| 2OIU | 60 | 63 | 62 | 62 | 64 |
| 2YGH | 37 | 37 | 37 | 39 | 43 |
| 3DIZ | 88 | 89 | 89 | 89 | 88 |
| 3E5E | 20 | 20 | 20 | 20 | 21 |
| 2GIS | 40 | 36 | 38 | 37 | 40 |
| 2CKY | 70 | 67 | 67 | 66 | 69 |
| 2PN3 | 16 | 17 | 16 | 16 | 16 |
| 3F2Q | 43 | 44 | 44 | 44 | 44 |
| 3IWN | 64 | 64 | 65 | 65 | 77 |
| 3BO3 | 86 | 85 | 81 | 85 | 92 |
| 3R4F | 25 | 21 | 21 | 21 | 23 |
| 3P49 | 44 | 43 | 46 | 44 | 60 |
| 1Y0Q | 97 | 89 | 87 | 92 | 104 |
| 3OTO | 561 | 593 | 596 | 587 | 692 |
| Average (%)[a] | 83.4 | 82.7 | 82.7 | 82.8 | 87.0 |
| Avg. high-res[b] | 91.4 | 92.0 | 92.0 | 92.7 | 92.4 |
| Avg. low-res[c] | 72.2 | 69.4 | 69.3 | 70.1 | 81.5 |
| Equal to or better than PDB | | 16 / 24 | 16 / 24 | 18 / 24 | 21 / 24 |

[a] Average value of (# of base-pairs) / (# of residues) $\times$ 2. The normalization is based on that the # of base-pairs in an ideal RNA duplex is half of the # of residues.
[b] Average value calculated for the five high resolution models 3DIL, 1U8D, 3D2V, 2GDI and 3TZR. Ultra-high resolution dataset 2A43 was excluded.
[c] Average value for the six lowest resolution models 3IWN, 3BO3, 3R4F, 3P49, 1Y0Q and 3OTO.

**Supplementary Table 6.** Summary of notable changes of base orientations (χ angle).

| PDB ID | Resolution | Chain-Residue | Base Type | PDB syn/anti | PDB χ | ERRASER-PHENIX syn/anti | ERRASER-PHENIX χ | Reference PDB | Resolution | Chain-Residue | PDB syn/anti | PDB χ |
|---|---|---|---|---|---|---|---|---|---|---|---|---|
| 3DIL[a] | 1.9 | A-110 | G | syn | 75 | anti | -88 | NA | | | | |
| 3D2V | 2 | B-35 | U | syn | 44 | anti | -123 | 3D2V | 2 | A-35 | anti | -110 |
| 2GDI | 2.05 | X-55 | C | anti | -65 | anti | -160 | NA | | | | |
| | | Y-55 | C | syn | 52 | anti | -135 | | | | | |
| 2QUS[b] | 2.4 | A-22 | G | syn | 97 | anti | -78 | NA | | | | |
| | | B-23 | C | syn | 56 | anti | -119 | | | | | |
| 2OIU | 2.6 | P-19 | U | syn | 58 | anti | -136 | 2OIU | 2.6 | Q-19 | anti | -112 |
| 2YGH | 2.6 | A-14 | A | syn | 41 | anti | -148 | 3GX5 | 2.4 | A-9 | anti | -85 |
| 3DIZ[a] | 2.85 | A-110 | G | syn | 80 | anti | -85 | NA | | | | |
| 2CKY | 2.9 | A-1 | G | syn | -7 | anti | 174 | 3D2V | 2 | A-1 | anti | -178 |
| | | A-35 | U | syn | 51 | anti | -116 | | | A-35 | anti | -110 |
| | | A-74 | U | syn | 36 | anti | -167 | | | A-74 | anti | -113 |
| | | B-67 | C | syn | -39 | anti | -129 | | | B-67 | anti | -100 |
| | | B-74 | U | anti | -50 | anti | -140 | | | B-74 | anti | -156 |
| 3IWN | 3.2 | A-7 | C | anti | -51 | anti | -162 | 3MXH | 2.3 | R-17 | anti | -155 |
| | | A-33 | A | anti | -100 | syn | 82 | | | R-33 | syn | 67 |
| 3BO3 | 3.4 | B-21 | C | syn | 91 | anti | -150 | NA | | | | |
| 3P49 | 3.55 | A-731 | C | anti | -58 | anti | -152 | NA | | | | |
| | | A-732 | C | syn | -33 | anti | -129 | | | | | |
| 1Y0Q[c] | 3.6 | A-35 | A | anti | 150 | anti | -77 | 3BO3 | 3.4 | NA | | |
| | | A-113 | A | syn | 21 | anti | -176 | | | B-121 | anti | 177 |
| | | A-158 | A | syn | 51 | anti | -152 | | | B-134 | anti | -167 |
| 3OTO | 3.69 | A-91 | C | anti | -58 | anti | -164 | 2VQE | 2.5 | A-91 | anti | -161 |
| | | A-108 | G | syn | -11 | anti | 180 | | | A-108 | syn | 3 |
| | | A-266 | G | anti | -59 | syn | 54 | | | A-266 | anti | -83 |
| | | A-328 | C | syn | 116 | anti | -85 | | | A-328 | syn | 110 |
| | | A-346 | G | syn | 46 | anti | -143 | | | A-346 | syn | 61 |
| | | A-365 | U | syn | 43 | anti | -157 | | | A-365 | syn | 53 |
| | | A-421 | U | syn | 33 | anti | -134 | | | A-421 | syn | 6 |
| | | A-839 | U | syn | 72 | anti | -164 | | | A-839 | syn | 117 |
| | | A-1004 | A | anti | -87 | syn | 67 | | | A-1004 | anti | -67 |
| | | A-1054 | C | syn | 81 | anti | -139 | | | A-1054 | syn | 77 |
| | | A-1279 | A | syn | 43 | anti | -101 | | | A-1279 | syn | 83 |

The table shows all the residues with Δχ > 90º upon ERRASER-PHENIX refinement. The definition of syn and anti conformation is: syn: -40 < χ ≤ 140; otherwise is anti. See supplementary results for discussion.

[a] 3DIL and 3DIZ are structures of the same sequence with different resolution. However, ERRASER flipped residue 110 in both structure from syn to anti. Therefore we did not perform the high-resolution vs. low-resolution comparison in the table.

[b] For 2QUS, the backbone conformations for residue 22-23 for chain A and chain B, as well as the structures of nearby region, are quite different. Therefore we did not use the different chains as reference models in our analysis.

[c] 1Y0Q was compared to a homologous structure 3BO3. The two structures are group I ribozymes from different species. Residue 35 has no homologous partner in 3BO3 so was not compared.

**Supplementary Table 7.** *R* factors of the benchmark.

| PDB ID | PDB (Deposited) | PDB (PHENIX calculated) | PHENIX | RNABC-PHENIX | RCrane-PHENIX | ERRASER-PHENIX |
|---|---|---|---|---|---|---|
| 2A43 | 0.114 | 0.128 | 0.140 | 0.140 | 0.145 | 0.137 |
| 3DIL | 0.192 | 0.183 | 0.172 | 0.169 | 0.171 | 0.170 |
| 1U8D | 0.178 | 0.177 | 0.166 | 0.167 | 0.165 | 0.162 |
| 3D2V | 0.207 | 0.202 | 0.212 | 0.210 | 0.224 | 0.211 |
| 2GDI | 0.208 | 0.199 | 0.188 | 0.189 | 0.196 | 0.187 |
| 3TZR | 0.182 | 0.182 | 0.184 | 0.185 | 0.188 | 0.180 |
| 3MXH | 0.202 | 0.222 | 0.196 | 0.196 | 0.202 | 0.199 |
| 2PN4 | 0.261 | 0.258 | 0.264 | 0.270 | 0.268 | 0.264 |
| 2QUS | 0.184 | 0.184 | 0.190 | 0.188 | 0.189 | 0.186 |
| 1Y27 | 0.232 | 0.224 | 0.204 | 0.205 | 0.207 | 0.203 |
| 2OIU | 0.203 | 0.196 | 0.190 | 0.193 | 0.196 | 0.194 |
| 2YGH | 0.200 | 0.186 | 0.197 | 0.206 | 0.199 | 0.205 |
| 3DIZ | 0.193 | 0.200 | 0.198 | 0.201 | 0.203 | 0.202 |
| 3E5E | 0.222 | 0.224 | 0.201 | 0.199 | 0.270 | 0.196 |
| 2GIS | 0.266 | 0.249 | 0.216 | 0.219 | 0.218 | 0.216 |
| 2CKY | 0.183 | 0.194 | 0.180 | 0.179 | 0.175 | 0.176 |
| 2PN3 | 0.229 | 0.214 | 0.213 | 0.212 | 0.239 | 0.210 |
| 3F2Q | 0.200 | 0.197 | 0.199 | 0.202 | 0.202 | 0.200 |
| 3IWN | 0.222 | 0.218 | 0.224 | 0.224 | 0.239 | 0.215 |
| 3BO3 | 0.282 | 0.266 | 0.244 | 0.240 | 0.274 | 0.239 |
| 3R4F | 0.239 | 0.221 | 0.184 | 0.182 | 0.187 | 0.179 |
| 3P49 | 0.282 | 0.279 | 0.227 | 0.233 | 0.218 | 0.233 |
| 1Y0Q[a] | 0.277 | 0.265 | 0.227 | 0.221 | 0.225 | 0.226 |
| 3OTO | 0.173 | 0.177 | 0.163 | 0.164 | 0.163 | 0.186 |
| Average | 0.214 | 0.210 | 0.199 | 0.200 | 0.207 | 0.199 |
| Avg. high-res[a] | 0.193 | 0.188 | 0.185 | 0.184 | 0.189 | 0.182 |
| Avg. low-res[b] | 0.246 | 0.238 | 0.211 | 0.211 | 0.218 | 0.213 |
| Equal to or better than PDB | | | 16 / 24 | 15 / 24 | 12 / 24 | 16 / 24 |

[a] Average value for the five high resolution models 3DIL, 1U8D, 3D2V, 2GDI and 3TZR. Ultra-high resolution dataset 2A43 was excluded.
[b] Average value for the six lowest resolution models 3IWN, 3BO3, 3R4F, 3P49, 1Y0Q and 3OTO.

**Supplementary Table 8.** $R_{\text{free}}$ factors of the benchmark.

| PDB ID | PDB (Deposited) | PDB (PHENIX calculated) | PHENIX | RNABC-PHENIX | RCrane-PHENIX | ERRASER-PHENIX |
|---|---|---|---|---|---|---|
| 2A43 | 0.190 | 0.180 | 0.166 | 0.165 | 0.176 | 0.162 |
| 3DIL | 0.229 | 0.213 | 0.194 | 0.193 | 0.199 | 0.199 |
| 1U8D | 0.228 | 0.218 | 0.198 | 0.202 | 0.202 | 0.198 |
| 3D2V | 0.251 | 0.244 | 0.233 | 0.235 | 0.249 | 0.239 |
| 2GDI | 0.241 | 0.229 | 0.216 | 0.219 | 0.219 | 0.217 |
| 3TZR | 0.245 | 0.242 | 0.239 | 0.239 | 0.246 | 0.240 |
| 3MXH | 0.239 | 0.270 | 0.231 | 0.233 | 0.239 | 0.235 |
| 2PN4 | 0.320 | 0.323 | 0.320 | 0.319 | 0.323 | 0.320 |
| 2QUS | 0.253 | 0.253 | 0.235 | 0.231 | 0.234 | 0.225 |
| 1Y27 | 0.264 | 0.255 | 0.235 | 0.238 | 0.238 | 0.235 |
| 2OIU | 0.238 | 0.233 | 0.224 | 0.225 | 0.225 | 0.224 |
| 2YGH | 0.259 | 0.244 | 0.234 | 0.234 | 0.237 | 0.236 |
| 3DIZ | 0.244 | 0.246 | 0.239 | 0.239 | 0.240 | 0.235 |
| 3E5E | 0.259 | 0.258 | 0.254 | 0.256 | 0.348 | 0.261 |
| 2GIS | 0.289 | 0.270 | 0.252 | 0.252 | 0.256 | 0.246 |
| 2CKY | 0.250 | 0.253 | 0.236 | 0.235 | 0.231 | 0.234 |
| 2PN3 | 0.283 | 0.274 | 0.272 | 0.273 | 0.282 | 0.267 |
| 3F2Q | 0.243 | 0.254 | 0.249 | 0.255 | 0.256 | 0.260 |
| 3IWN | 0.292 | 0.287 | 0.282 | 0.281 | 0.297 | 0.270 |
| 3BO3 | 0.325 | 0.312 | 0.295 | 0.293 | 0.316 | 0.295 |
| 3R4F | 0.271 | 0.252 | 0.245 | 0.241 | 0.243 | 0.239 |
| 3P49 | 0.310 | 0.299 | 0.290 | 0.279 | 0.280 | 0.295 |
| 1Y0Q | 0.310 | 0.307 | 0.294 | 0.289 | 0.292 | 0.291 |
| 3OTO | 0.231 | 0.232 | 0.225 | 0.225 | 0.223 | 0.233 |
| Average | 0.261 | 0.256 | 0.244 | 0.244 | 0.252 | 0.244 |
| Avg. high-res[a] | 0.239 | 0.229 | 0.216 | 0.218 | 0.223 | 0.219 |
| Avg. low-res[b] | 0.290 | 0.281 | 0.272 | 0.268 | 0.275 | 0.270 |
| Equal to or better than PDB | | | 24 / 24 | 24 / 24 | 17 / 24 | 21 / 24 |

[a] Average value for the five high resolution models 3DIL, 1U8D, 3D2V, 2GDI and 3TZR. Ultra-high resolution dataset 2A43 was excluded.
[b] Average value for the six lowest resolution models 3IWN, 3BO3, 3R4F, 3P49, 1Y0Q and 3OTO.

**Supplementary Table 9.** Similarity analysis between high-resolution and low-resolution models.

| High res. models | Low res. models | Similar residues (%)[a] | | | | | Similar puckers (%)[b] | | | | |
|---|---|---|---|---|---|---|---|---|---|---|---|
| | | PDB | PHENIX | RNABC-PHENIX | RCrane-PHENIX | ERRASER-PHENIX | PDB | PHENIX | RNABC-PHENIX | RCrane-PHENIX | ERRASER-PHENIX |
| 4FE5[c] | 1U8D | 87.5 | 87.5 | 87.5 | 89.1 | 92.2 | 98.4 | 96.9 | 96.9 | 98.4 | 100.0 |
| 4FE5[c] | 1Y27 | 67.2 | 78.1 | 78.1 | 82.8 | 85.9 | 93.8 | 100.0 | 100.0 | 100.0 | 100.0 |
| 1U8D | 1Y27 | 68.8 | 76.6 | 75.0 | 84.4 | 82.8 | 95.3 | 96.9 | 96.9 | 98.4 | 96.9 |
| 2YGH | 2GIS | 65.6 | 66.7 | 61.3 | 68.8 | 81.7 | 88.2 | 92.5 | 92.5 | 95.7 | 97.9 |
| 3DIL | 3DIZ | 94.3 | 93.7 | 92.5 | 85.1 | 94.8 | 98.9 | 98.9 | 98.9 | 98.9 | 98.3 |
| 3MXH | 3IWN_1 | 53.3 | 54.6 | 55.8 | 61.0 | 71.4 | 85.7 | 92.2 | 92.2 | 89.6 | 94.8 |
| 3MXH | 3IWN_2 | 45.5 | 49.4 | 48.1 | 50.7 | 63.6 | 81.8 | 90.9 | 90.9 | 89.6 | 90.9 |
| 3D2V_1 | 2CKY_1 | 57.1 | 71.4 | 72.7 | 66.2 | 79.2 | 85.7 | 96.1 | 96.1 | 94.8 | 97.4 |
| 3D2V_1 | 2CKY_2 | 42.9 | 64.9 | 67.5 | 74.0 | 75.3 | 89.6 | 97.4 | 97.4 | 97.4 | 94.8 |
| 3D2V_2 | 2CKY_1 | 54.6 | 61.0 | 62.3 | 63.6 | 77.9 | 92.2 | 100.0 | 100.0 | 98.7 | 100.0 |
| 3D2V_2 | 2CKY_2 | 45.5 | 59.7 | 64.9 | 68.8 | 71.4 | 90.9 | 98.7 | 97.4 | 97.4 | 97.4 |
| 2PN4_1 | 2PN3 | 77.3 | 84.1 | 81.8 | 86.4 | 86.4 | 95.5 | 97.7 | 97.7 | 93.2 | 97.7 |
| 2PN4_2 | 2PN3 | 84.1 | 84.1 | 86.4 | 81.8 | 84.1 | 93.2 | 95.5 | 95.5 | 93.2 | 95.5 |
| Average | | 64.9 | 71.7 | 71.9 | 74.1 | 80.5 | 91.5 | 96.4 | 96.3 | 95.8 | 97.0 |
| Equal to or better than PDB | | | 12 / 13 | 11 / 13 | 11 / 13 | 13 / 13 | | 12 / 13 | 12 / 13 | 12 / 13 | 12 / 13 |

[a] Nucleotide pair in which the differences between all torsion angles are smaller than 40°.
[b] Nucleotide pair in which the difference between δ angle is smaller than 20°.
[c] 4FE5 is a ultra-high resolution (1.32 Å) guanine riboswitch structure not in the benchmark set.

**Supplementary Table 10.** Torsional RMSDs (in degrees) between high-resolution and low-resolution models.

| High res. models | Low res. models | PDB | PHENIX | RNABC-PHENIX | RCrane-PHENIX | ERRASER-PHENIX |
|---|---|---|---|---|---|---|
| 4FE5 | 1U8D | 25.1 | 24.5 | 25.8 | 21.5 | 15.3 |
| 4FE5 | 1Y27 | 39.3 | 36.7 | 33.5 | 33.2 | 30.5 |
| 1U8D | 1Y27 | 37.3 | 35.1 | 32.0 | 30.3 | 31.4 |
| 2YGH | 2GIS | 38.4 | 37.7 | 38.2 | 32.2 | 29.0 |
| 3DIL | 3DIZ | 13.5 | 12.8 | 13.4 | 26.2 | 16.4 |
| 3MXH | 3IWN_1 | 48.0 | 45.6 | 44.2 | 41.3 | 37.3 |
| 3MXH | 3IWN_2 | 51.9 | 49.5 | 48.8 | 48.2 | 42.2 |
| 3D2V_1 | 2CKY_1 | 41.0 | 38.7 | 37.5 | 38.7 | 33.0 |
| 3D2V_1 | 2CKY_2 | 40.9 | 37.4 | 35.1 | 33.8 | 35.2 |
| 3D2V_2 | 2CKY_1 | 42.6 | 39.5 | 38.6 | 38.8 | 31.5 |
| 3D2V_2 | 2CKY_2 | 42.7 | 40.2 | 39.7 | 37.2 | 36.9 |
| 2PN4_1 | 2PN3 | 25.0 | 21.7 | 21.8 | 23.6 | 23.8 |
| 2PN4_2 | 2PN3 | 24.5 | 22.9 | 22.7 | 22.8 | 23.9 |
| Average | | 36.2 | 34.0 | 33.2 | 32.9 | 29.7 |
| Equal to or better than PDB | | | 13 / 13 | 12 / 13 | 12 / 13 | 12 / 13 |

RMSD is calculated between all the torsion angles in the model pair.

**Supplementary Table 11.** Similarity analysis for model pairs of the same or similar sequences.

| Chain1 | Chain 2 | Similar residues (%)[a] | | | | | Similar puckers (%)[b] | | | | |
|---|---|---|---|---|---|---|---|---|---|---|---|
| | | PDB | PHENIX | RNABC-PHENIX | RCrane-PHENIX | ERRASER-PHENIX | PDB | PHENIX | RNABC-PHENIX | RCrane-PHENIX | ERRASER-PHENIX |
| 2GDI_1 | 2GDI_2 | 78.8 | 86.1 | 86.1 | 86.1 | 86.1 | 97.5 | 100.0 | 100.0 | 98.7 | 98.7 |
| 2OIU_1 | 2OIU_2 | 42.3 | 53.5 | 57.8 | 63.4 | 69.0 | 87.3 | 94.4 | 94.4 | 94.4 | 95.8 |
| 2QUS_1 | 2QUS_2 | 60.9 | 75.4 | 75.4 | 75.4 | 81.2 | 92.8 | 94.2 | 94.2 | 94.2 | 94.2 |
| 3P49_1 | 3P49_2 | 50.7 | 50.7 | 53.3 | 58.4 | 68.8 | 94.8 | 97.4 | 97.4 | 97.4 | 92.2 |
| 1U8D | 1Y27 | 77.9 | 77.9 | 80.5 | 88.3 | 87.0 | 97.4 | 98.7 | 98.7 | 97.4 | 97.4 |
| 2YGH | 2GIS | 59.7 | 71.4 | 72.7 | 84.4 | 84.4 | 90.9 | 98.7 | 98.7 | 98.7 | 98.7 |
| 3DIL | 3DIZ | 86.4 | 90.9 | 88.6 | 90.9 | 90.9 | 97.7 | 97.7 | 97.7 | 97.7 | 97.7 |
| 3MXH | 3IWN_1 | 34.2 | 39.5 | 34.2 | 68.4 | 73.7 | 84.2 | 94.7 | 92.1 | 94.7 | 100.0 |
| 3MXH | 3IWN_2 | 68.8 | 78.1 | 82.8 | 85.9 | 84.4 | 95.3 | 100.0 | 100.0 | 100.0 | 100.0 |
| 3IWN_1 | 3IWN_2 | 65.6 | 71.0 | 66.7 | 71.0 | 89.3 | 88.2 | 93.6 | 93.6 | 95.7 | 97.9 |
| 3D2V_1 | 3D2V_2 | 94.3 | 95.4 | 94.8 | 92.5 | 97.7 | 98.9 | 100.0 | 100.0 | 100.0 | 100.0 |
| 3D2V_1 | 2CKY_1 | 53.3 | 55.8 | 54.6 | 64.9 | 72.7 | 85.7 | 92.2 | 92.2 | 92.2 | 96.1 |
| 3D2V_1 | 2CKY_2 | 45.5 | 48.1 | 49.4 | 55.8 | 62.3 | 81.8 | 92.2 | 92.2 | 93.5 | 90.9 |
| 3D2V_2 | 2CKY_1 | 57.1 | 70.1 | 70.1 | 77.9 | 85.7 | 85.7 | 98.7 | 98.7 | 98.7 | 98.7 |
| 3D2V_2 | 2CKY_2 | 42.9 | 64.9 | 70.1 | 79.2 | 77.9 | 89.6 | 100.0 | 100.0 | 100.0 | 97.4 |
| 2CKY_1 | 2CKY_2 | 54.6 | 63.6 | 64.9 | 79.2 | 83.1 | 92.2 | 100.0 | 100.0 | 98.7 | 98.7 |
| 2PN4_1 | 2PN4_2 | 45.5 | 62.3 | 64.9 | 77.9 | 74.0 | 90.9 | 98.7 | 98.7 | 97.4 | 100.0 |
| 2PN4_1 | 2PN3 | 77.3 | 81.8 | 79.6 | 84.1 | 81.8 | 95.5 | 97.7 | 97.7 | 93.2 | 97.7 |
| 2PN4_2 | 2PN3 | 84.1 | 84.1 | 86.4 | 93.2 | 86.4 | 93.2 | 95.5 | 95.5 | 95.5 | 95.5 |
| Average | | 62.1 | 69.5 | 70.2 | 77.7 | 80.9 | 91.6 | 97.1 | 96.9 | 96.7 | 97.2 |
| Equal to or better than PDB | | | 19 / 19 | 19 / 19 | 18 / 19 | 19 / 19 | | 19 / 19 | 19 / 19 | 18 / 19 | 18 / 19 |

[a] Nucleotide pair in which the differences between all torsion angles are smaller than 40°.
[b] Nucleotide pair in which the difference between δ angle is smaller than 20°.

**Supplementary Table 12.** Torsional RMSDs (in degrees) for model pairs of the same or similar sequences.

| Chain 1 | Chain 2 | PDB | PHENIX | RNABC-PHENIX | RCrane-PHENIX | ERRASER-PHENIX |
|---|---|---|---|---|---|---|
| 2GDI_1 | 2GDI_2 | 25.0 | 23.4 | 23.4 | 24.3 | 22.8 |
| 2OIU_1 | 2OIU_2 | 44.5 | 41.7 | 39.8 | 39.9 | 34.6 |
| 2QUS_1 | 2QUS_2 | 46.6 | 42.3 | 40.4 | 32.5 | 29.0 |
| 3P49_1 | 3P49_2 | 42.5 | 41.6 | 40.8 | 36.1 | 38.5 |
| 1U8D | 1Y27 | 31.6 | 31.4 | 25.3 | 19.4 | 22.2 |
| 2YGH | 2GIS | 32.1 | 31.9 | 32.0 | 24.5 | 23.2 |
| 3DIL | 3DIZ | 19.2 | 17.5 | 22.4 | 17.7 | 16.4 |
| 3MXH | 3IWN_1 | 50.9 | 49.7 | 51.7 | 37.0 | 32.9 |
| 3MXH | 3IWN_2 | 37.3 | 34.4 | 26.0 | 28.2 | 28.8 |
| 3IWN_1 | 3IWN_2 | 38.4 | 37.4 | 37.5 | 30.5 | 22.1 |
| 3D2V_1 | 3D2V_2 | 13.5 | 12.0 | 11.7 | 19.5 | 11.7 |
| 3D2V_1 | 2CKY_1 | 48.0 | 45.6 | 44.0 | 38.0 | 37.3 |
| 3D2V_1 | 2CKY_2 | 51.9 | 49.4 | 46.7 | 42.1 | 42.5 |
| 3D2V_2 | 2CKY_1 | 41.0 | 38.5 | 35.1 | 31.4 | 27.6 |
| 3D2V_2 | 2CKY_2 | 40.9 | 36.9 | 30.8 | 27.3 | 31.9 |
| 2CKY_1 | 2CKY_2 | 42.6 | 39.6 | 37.8 | 27.5 | 24.9 |
| 2PN4_1 | 2PN4_2 | 42.7 | 40.3 | 37.8 | 27.5 | 31.0 |
| 2PN4_1 | 2PN3 | 25.0 | 21.2 | 26.7 | 24.1 | 22.2 |
| 2PN4_2 | 2PN3 | 24.5 | 21.3 | 20.0 | 16.5 | 20.2 |
| Average | | 36.7 | 34.5 | 33.2 | 28.6 | 27.4 |
| Equal to or better than PDB | | | 19 / 19 | 16 / 19 | 18 / 19 | 19 / 19 |

RMSD is calculated between all the torsion angles in the model pair.

**Supplementary Table 13.** P-values of Wilcoxon signed-rank test between each method and the starting PDB dataset for all geometric features, $R$ and $R_{\text{free}}$.

| | Outlier bond | Outlier angle | Clashscore | Outlier backbone rotamer | Potentially incorrect pucker | Number of base-pairs | $R$ | $R_{\text{free}}$ |
|---|---|---|---|---|---|---|---|---|
| PHENIX | 0.017 | 0.004 | 0.089 | 0.001 | < 0.001 | 0.521 | 0.009 | < 0.001 |
| RNABC -PHENIX | 0.017 | 0.005 | 0.045 | 0.007 | < 0.001 | 0.394 | 0.024 | < 0.001 |
| RCrane -PHENIX | 0.017 | 0.015 | 0.136 | < 0.001 | < 0.001 | 0.733 | 0.450 | 0.014 |
| ERRASER -PHENIX | 0.018 | 0.005 | 0.006 | < 0.001 | < 0.001 | 0.009 | 0.011 | < 0.001 |

The Wilcocon signed-rank test is a non-parametric statistical test for pairs of related samples. The test can tell whether two datasets are significantly different from each other. Therefore it is suitable for the comparison between the PDB-deposited values and values after improvement methods. Here the two-sided P-value for each comparison is given in the table. All data in the benchmark ($n$ = 24) are used for the analysis. Calculations are performed with the python library SciPy.

# Supplementary Notes

*Example Rosetta command lines used in ERRASER-PHENIX*

(1) Full structure minimization.

```
erraser_minimizer.<exe> -native <input pdb> -out_pdb <output
pdb> -score::weights rna/rna_hires_elec_dens -
score:rna_torsion_potential RNA11_based_new –constrain_P true -
vary_geometry true -fixed_res <fixed residue list> -
edensity:mapfile <map file> -edensity:mapreso 2.0 -
edensity:realign no
```

(2) Single nucleotide rebuilding with analytical chain closure.

```
swa_rna_analytical_closure.<exe> -algorithm rna_resample_test -s
<input pdb> -native <native pdb> -out:file:silent blah.out -
sampler_extra_syn_chi_rotamer true -
sampler_extra_anti_chi_rotamer true -constraint_chi true -
sampler_cluster_rmsd 0.1 -sampler_native_rmsd_screen true -
sampler_native_screen_rmsd_cutoff 3.0 -sampler_num_pose_kept 100
-PBP_clustering_at_chain_closure true -
allow_chain_boundary_jump_partner_right_at_fixed_BP true -
add_virt_root true - rm_virt_phosphate true -sample_res 2 -
cutpoint_closed 2  -fasta fasta -input_res 1 3-4 -fixed_res 1 3-
4 -jump_point_pairs NOT_ASSERT_IN_FIXED_RES 1-4 -alignment_res
1-4 -rmsd_res 4 -score:weights rna/rna_hires_elec_dens -
edensity:mapfile <map file> -edensity:mapreso 2.0 -
edensity:realign no -score:rna_torsion_potential RNA11_based_new
```

(3) Single nucleotide rebuild at terminal nucleotides.

```
swa_rna_main.<exe> -algorithm rna_resample_test -s <input pdb> -
native <native pdb> -out:file:silent blah.out -
sampler_extra_syn_chi_rotamer true -
sampler_extra_anti_chi_rotamer true -constraint_chi true -
sampler_cluster_rmsd 0.1 -sampler_native_rmsd_screen true -
sampler_native_screen_rmsd_cutoff 3.0 -sampler_num_pose_kept 100
-PBP_clustering_at_chain_closure true -
allow_chain_boundary_jump_partner_right_at_fixed_BP true -
add_virt_root true - rm_virt_phosphate true -sample_res 2 -
cutpoint_closed 2 -fasta fasta -input_res 1-4 -fixed_res 2-4 -
jump_point_pairs NOT_ASSERT_IN_FIXED_RES 1-4 -alignment_res 1-4
```

```
-rmsd_res 4 -score:weights rna/rna_hires_elec_dens -
edensity:mapfile <map file> -edensity:mapreso 2.0 -
edensity:realign no -score:rna_torsion_potential RNA11_based_new
```

*Example PHENIX command lines used in ERRASER-PHENIX*

(1) phenix.ready_set.

```
phenix.ready_set 3E5E.pdb
```

(2) One-cycle TLS refinement.

```
phenix.refine 2QUS-sf.mtz 2QUS.updated.pdb GTP.cif
2QUS.metal.edits 2QUS.link.edits tls.params
main.number_of_macro_cycles=1 strategy=tls
```

(3) Three-cycle refinement with manual parameter set.

```
phenix.refine 2GIS-sf.mtz  2GIS.pdb  2GDI.metal.edits
ordered_solvent=true  optimize_adp_weight=true
strategy=individual_sites+individual_adp+occupancies
main.number_of_macro_cycles=3 wxc_scale=0.1
```

(4) Three-cycle refinement with automatically optimized target weight.

```
phenix.refine 2GIS-sf.mtz
2GIS_phenix_erraser_refine_001_refine_001.pdb
2GIS_phenix_erraser_refine_001_refine_001.metal.edits
ordered_solvent=true optimize_adp_weight=true
strategy=individual_sites+individual_adp+occupancies
main.number_of_macro_cycles=3 optimize_xyz_weight=true
```

(5) Density map creation.

```
phenix.maps maps.params

maps.params:
...
   map {
     map_type = 2mFo-DFc
     format = xplor *ccp4
     file_name = 2GIS_cell.ccp4
```

```
        kicked = True
        fill_missing_f_obs = True
        grid_resolution_factor = 1/4.
        region = selection *cell
        atom_selection = None
        atom_selection_buffer = 3
        sharpening = False
        sharpening_b_factor = None
        exclude_free_r_reflections = True
        isotropize = True
    }
...
```

## Supplementary Results

*ERRASER-PHENIX improves RNA base-pairing geometry*

ERRASER-PHENIX visually improved the base pairing patterns of the RNA models, enhancing the co-planarity of interacting bases. For example, Figure 1f shows a helical region in 3P49, a structure solved at 3.55 Å resolution. At this resolution, accurate positioning of base planes into the electron density map was difficult. Manual fits gave base pairs that were buckled or twisted compared to geometries seen in higher-resolution crystallographic models[3]. RNABC and RCrane held the base positions fixed during rebuilding and were thus unable to improve the base-pair planarity. On the other hand, ERRASER-PHENIX improved the planarity of the base-pairs, likely due to the hydrogen bonding potential included in the Rosetta energy function. Independent base-pair validation tools – which, like MolProbity, would permit unbiased assessment of improvement – are not currently available. However, we applied the base-pair assignment method MC-Annotate[4] and noted that the refined structures gave a higher number of automatically assigned base-pairs than the starting PDB models in 21 out of 24 cases (Supplementary Table 5). For the 3P49 case, ERRASER-PHENIX increased the number of base-pairs from 44 in the PDB model to 60. Other methods (RNABC-PHENIX and RCrane-PHENIX) lead to smaller improvements in this case, giving 46 and 44 base-pairs respectively.

*ERRASER-PHENIX improves the base orientation*

In addition to the base-pairing geometry improvement described above, ERRASER also improved the orientation of bases in the models. The glycosidic torsions in a RNA structure predominantly adopt two distinct conformations: syn and anti[5]. These two conformations can be distinguished by the value of $\chi$ torsion angle of the glycosidic bond. In the discussion below, the syn conformation is defined as $-40º < \chi \leq 140º$, and the remaining angle ranges are defined as anti, based on the distribution of $\chi$ angles in RNA09 dataset (Supplementary Fig. 2a, http://kinemage.biochem.duke.edu/databases/rnadb.php). It is also evident that the anti conformation is much more probable than syn conformation, and syn pyrimidine conformers are especially rare. Therefore in the current single-nucleotide rebuilding scheme, syn pyrimidine conformers were sampled only if the starting model adopts syn conformation. Supplementary Table 6 summarized all the notable base orientation changes ($\Delta\chi > 90º$) in the benchmark. While most of the changes are syn-to-anti flips, in agreement with the higher frequency of anti conformations, there are still a few anti-to-syn flips, confirming that ERRASER did not blindly flip the base orientation from syn to anti. To evaluate the confidence of these remodeled bases, these changes were compared to reference models of similar or higher resolution with same or similar sequence. If such models were not available, and there were two different copies of structures in one asymmetric unit, the different copy (to which the target nucleotide did not belong) was used as reference model. For all test cases where reference models are available except for the ribosome, all 12 base orientation changes agreed well with the reference. In some cases these orientation changes even introduced extra hydrogen bonds, further supporting these fixes (Supplementary Fig. 2b). For the lowest-resolution ribosome test case (3OTO, 3.69 Å), most of the conformational changes did not match the higher-resolution reference model (2VGE, 2.5 Å). However, since both structures were solved using molecular replacement, it is possible that the two deposited structures inherited the same base orientations from an earlier model[6-7], therefore might share the same erroneous conformations. By detailed inspection of the conformational changes, we found that the ERRASER changes in 3OTO gave the same or more hydrogen bonds as the starting coordinates and agreed well with the electron density. For example, Supplementary Figure 2c shows an example of such an orientation change where a guanosine flipped from anti to syn and forming a Watson-Crick vs. Hoogsten base-pair with two extra hydrogen bonds. The density map of higher-resolution model did not falsify the possibility of this new, energetically more favorable conformation. On the other hand, Supplementary Figure 2d demonstrates a case where the flipping is ambiguous, where a uridine flipped from syn to anti, and both conformations have one hydrogen bond with nearby nucleotide. Although the starting conformation fits slightly better in the higher-resolution density map visually, it is possible that the alternative conformation also existed in the crystal structure with a smaller occupancy. Across the benchmark, ERRASER-PHENIX gave improved, or at least alternatively possible, conformations for the base orientations in RNA.

*Pairwise comparison for models with similar sequences*

Analogous to the independent comparison between remodeled low-resolution and original high-resolution models (see the main text), we reasoned that pairs of models with the same or similar RNA sequences should give similar conformations at corresponding nucleotides, and an accurate refinement procedure should maintain or improve this similarity. We drew 19 such structural pairs from three categories: models of the same sequences determined independently at different resolutions (11 cases, same pairs as those used in high-res vs. low-res comparisons); two copies present in the asymmetric unit related by non-crystallographic symmetry (7 cases); and the conserved regions of two aptamer domains in glycine riboswitch (1 case). Supplementary Table 10-11 summarizes the results of the similarity comparison of the torsion angles of each nucleotide pair, sugar pucker assignment and the RMSD in torsional space for these structure pairs. In nearly all cases, ERRASER-PHENIX improved these metrics compared to the PDB models, and gave better average values than RNABC-PHENIX and RCrane-PHENIX.